\documentclass[12pt,preprint]{aastex}
\usepackage{natbib}
\begin{document}

\shorttitle{Heterogeneous Astrophysical Jets}
\shortauthors{Yirak et al.}

\title{Hypersonic Buckshot: Astrophysical Jets as Heterogeneous Collimated Plasmoids}
\author{Kristopher Yirak\altaffilmark{1}, Adam Frank\altaffilmark{1}, Andrew J. Cunningham\altaffilmark{1,2},Sorin Mitran\altaffilmark{3}}
\altaffiltext{1}{Department of Physics and Astronomy, University of Rochester, Rochester, NY 14620 \\Email contact: yirak@pas.rochester.edu}
\altaffiltext{2}{Lawrence Livermore National Laboratory, Livermore, CA 94551}
\altaffiltext{3}{Department of Mathematics, Applied Mathematics Program, University of North Carolina, Chapel Hill, NC 27599}

\begin{abstract}
Herbig-Haro (HH) jets are commonly thought of as homogeneous beams of plasma traveling at hypersonic velocities. Structure within jet beams is often attributed to periodic or ``pulsed'' variations of conditions at the jet source.  Simulations based on this scenario result in knots extending across the jet diameter. Observations and recent high energy density laboratory experiments shed new light on structures below this scale and indicate they may be important for understanding the fundamentals of jet dynamics. In this paper we offer an alternative to ``pulsed'' models of protostellar jets.  Using direct numerical simulations we explore the possibility that jets are chains of sub-radial clumps propagating through a moving inter-clump medium.  Our models explore an idealization of this scenario by injecting small ($r<r_{jet}$), dense ($\rho>\rho_{jet}$) spheres embedded in an otherwise smooth inter-clump jet flow. The spheres are initialized with velocities differing from the jet velocity by $\sim15$\%. We find the consequences of shifting from homogeneous to heterogeneous flows are significant as clumps interact with each other and with the inter-clump medium in a variety of ways.  Structures which mimic what is expected from pulsed-jet models can form, as can previously unseen ``sub-radial'' behaviors including backward facing bow shocks and off-axis working surfaces. While these small-scale structures have not been seen before in simulation studies, they are found in high resolution jet observations. We discuss implications of our simulations for the interpretation of protostellar jets with regard to characterization of knots by a ``lifetime'' or ``velocity history'' approach as well as linking observed structures with central engines which produce the jets.
\end{abstract}


\section{Introduction}
Herbig-Haro objects have been the subject of significant analytical, observational, and numerical attention since their discovery. Observations using optical \citep[e.g.][]{bally2002hh1/2} and IR \citep[e.g.][]{velusamy2007hh46/47} techniques reveal that these jets typically show striking large scale collimation extending out to parsec distances combined with features appearing on a range of smaller scales.

Structure along the jet beam (``knots'' or ``clumps'') have, in particular, received considerable attention.  The origin of knots remains a subject of debate. Early studies focused on clumpyness of the HH bow shocks; \cite{norman1979} postulated the existence of single ``interstellar bullets,'' while \cite{schwartz1978} attributed the structures to stationary clumps being overrun by a wind. Stationary crossing shocks due to an overpressured jet beam expanding and then re-collimating were an early possibility that was considered for knots along the beam \citep{buehrke1988,raga1990internal}. More recently, \cite{rubini2007obliqueshocks} have suggested oblique shock focusing as a natural mechanism for hydrodynamic knot formation, though the presence of magnetic fields \citep{hartigan2007bfields}, precession \citep{masciadri2002precession}, and interactions with the environment \citep{raga2002hh110, dalpino1999, yirak2008} all offer other means by which dense clumps might be created.  While considerable work has gone into these scenarios, currently the most favored model for the knots are internal working surfaces where shocks are driven down the beam by pulsation at the jet source. This ``pulsation'' model was first proposed by \cite{rees1978} and was extensively explored by Raga and collaborators \citep{raga1990ws, biro1994, raga1993}. In pulsed jet simulations the density and velocity cross-sectional profiles $\rho_j(r)$ \& $v_j(r)$ in the jet-launching region are kept fixed, while the magnitude of the velocity varies sinusoidally \citep{raga1990ws, volker1999}. The pulsation scenario has become so dominant that even when attempting to address questions unrelated to clump formation, periodic inflow variations are frequently employed \citep[e.g.][]{suttner1997}.

A variety of observational signatures can be recovered via pulsed jet models through careful choice of specific jet physical parameters and sinusoidal variability. In \cite{raga2002hh34/111}, for example, a two-mode launching model was proposed using velocity histories extracted from observations of HH~34 and HH~111. Using these pulsation modes axisymmetric hydrodynamic simulations provided a convincing match to the location of the leading bow shock and the location of bright knots in the beam. These and similar results provide strong support for pulsed jet models.

A detailed examination of jets observed at the highest spatial resolution however shows features which do not fit into the pulsed jet paradigm.  In particular a number of ``archetypal'' jets show features at scales below the jet radius ($r<r_j$) which are distinctly displaced from the jet axis. In the case of HH~47 the jet clearly shows a non-axisymmetric morphology in the form an apparent helical bending of the beam \citep{hartigan2005hh47}.  The beam itself is defined by a sequence of quasi-periodic knots with displacements to either side of the nominal jet axis. Explanations for this bending have included impacts with objects \citep{raga2002hh110, dalpino1999}, magnetic fields \citep{hartigan2007bfields} and precession of the jet source \citep{masciadri2002precession}. We consider the presence of sub-radial, non-axisymmetric features to be a challenge to the pulsed jet paradigm.  Here, we consider an alternative to the pulsed jet model and investigate the consequences of intrinsic density heterogeneity in jet formation and evolution.

We are motivated to explore this model both by observations and new high energy density laboratory astrophysics (HEDLA) plasma experiments.  Using pulsed power wire array technologies \cite{lebedev2005labjet}, \cite{ciardi2007labjet}, and Ciardi et al.~(2008, in preparation) have presented experiments that track the evolution of fully magnetized, hypersonic, radiative jets.  The stability of hydro and MHD jets has long been a topic of debate, and these experiments shed some light on the real dynamics of 3-D systems \citep{xu2000jetstability3d}. The experiments show that kink mode instabilities strongly affect the jet.  As the kink mode grows into the non-linear regime its disrupts but does not destroy the jet. The saturation of the instability transforms the jet into a sequence of collimated chains of knots which propagate with a range of velocities. Similar fragmented chains have been seen in other pulsed power experiments \citep{golingo2005zpinch}.

Thus we propose a model in which the velocity and density profile of the jet are variable in time and space on scales less than the jet age and jet radius, respectively. Specifying conditions on these scales allows the model to achieve complex structures not seen before in simulations. Our simulations utilize AMR techniques and so are able resolve clumps in the jet beam at acceptable levels.

In the following sections, we describe the model in detail, offer dynamical and observational signatures, and briefly discuss the results.

\section{Computational Method \& Physical Model}
Numerical simulations were undertaken with the \emph{AstroBEAR} computational code \footnote{Information about the \emph{AstroBEAR} code may be found online, at http://www.pas.rochester.edu/{\tiny $\sim$}bearclaw/}. $AstroBEAR$ is a parallel adaptive mesh refinement (AMR) code which allows a variety of choices for numerical solvers, integration schemes, and cooling modules for hydrodynamic or magnetohydrodynamic astrophysical fluids \citep{cunningham2006collisions,cunningham2007amrmhd}. Here, the code solved the 3D hyperbolic system of equations for inviscid, compressible flow using a spatial second-order and temporal first-order accurate MUSCL scheme using a Roe-averaged linearized Riemann solver. Simple radiative cooling is included separately using an iterative source term with the cooling curve of \cite{dalgarno1972}.

In what follows, the geometry of the system is taken to have the jet axis aligned with the $z$-axis, with $x$- and $y$-axes following the right handed convention. A base grid of $36\times36\times96$ cells covered the simulation domain of extent $1,200\times1,200\times3,200$ AU. AstroBEAR employs a patch-based adaptive grid to refine areas of interest; here, two levels of refinement were used, yielding a maximum effective resolution of $288\times288\times768$ cells.  This corresponds to a minimum cell length of $\Delta x=6.23\times10^{13}$ cm ($\sim$4 AU). All boundaries had outflow conditions, with user-specified conditions in the jet launching region.

The ambient number density and temperature were initialized to constant values of $\rho_a$=10 cm$^{-3}$ and $T_a$=2,000 K, respectively, giving a sound speed of $c_a=5.26$ km~s$^{-1}$. The inhomogeneities in the beam were introduced as spherical density and velocity perturbations in an otherwise smooth beam: we shall refer to the former as ``clumps'' and the latter as the ``jet.'' The jet had number density, radius, velocity, and temperature of $\rho_j$=10$^2$ cm$^{-3}$, $r_j$=100 AU, $v_j$=150 km~s$^{-1}$, and $T_j$=2,000 K. This resulted in a Mach 30, over-dense ($\chi_{ja}\equiv\rho_j/\rho_a=10$), and overpressured ($p_j/p_a=10$) jet. The jet was launched at $x,y$=600 AU on the $z=0$ plane. The clumps all had the same initial number density of $\rho_c$=10$^3$ cm$^{-3}$, yielding density ratios of $\chi_{cj}\equiv\rho_c/\rho_j$=10 and $\chi_{ca}\equiv\rho_c/\rho_a$=10$^2$. The clumps were seeded with radii and velocities that were random within ranges of $r_c$=26--60 AU, and $v_c$=132--168 km~s$^{-1}$. Relative to the jet, these correspond to ranges $r_c/r_j$=0.26--0.60 and $v_c/v_j$=0.88--1.12. The $x$- and $y$-locations of the clumps in the jet were random with the constraint that the entire clump be located within the jet beam. All clumps were seeded at the same  $z$ location in the grid. The clumps had temperatures such that they were in pressure balance with the jet beam (assuming a relative velocity of zero). They were seeded nominally 8 years apart; this resulted in the production of 12 clumps before the end of the simulation ($t_{sim}=100$ yr).

At the maximum AMR level, the jet radius was resolved by 24 cells, and the clump radii by 6--14 cells. This represents probably the lower limit on desirable resolution. Jet, clump, and ambient materials were separately tracked with passive advected tracers.

\section{Results \& Analysis}
\begin{figure}[htbp]
\centering
\includegraphics[width=120mm]{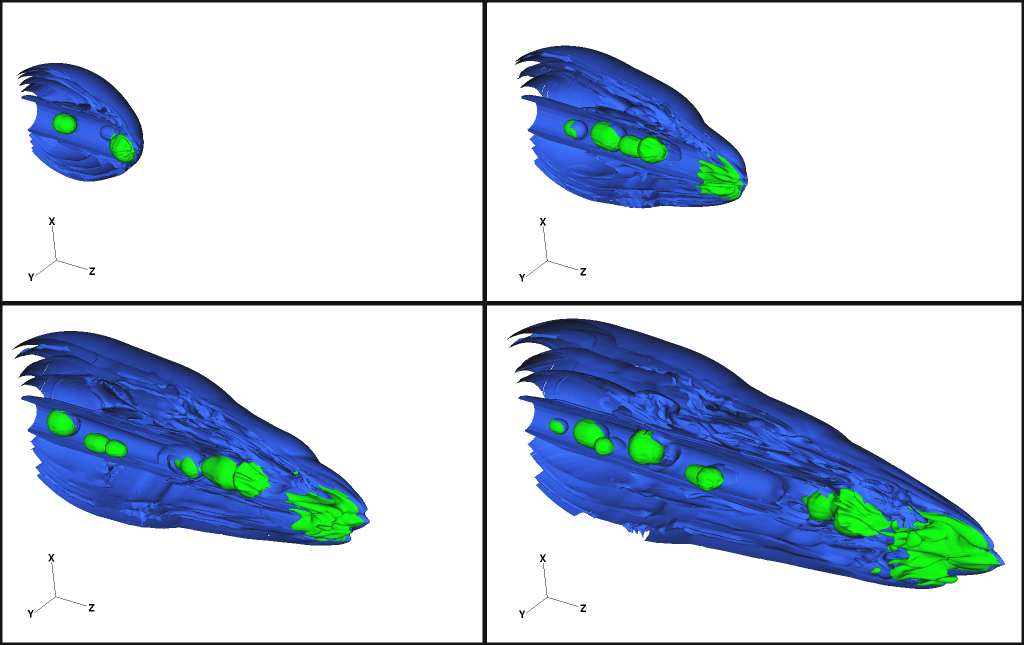}
\caption{Isocontours of logarithmic density at four times in the simulation, $t=$30, 53, 77, \& 100 yr. The clumps are depicted in light green, with the jet material in blue. The $x$-$z$ plane along the jet axis clips the jet material contours.}
\label{isocontours}
\end{figure}

\begin{figure}[htbp]
\centering
\includegraphics[width=120mm]{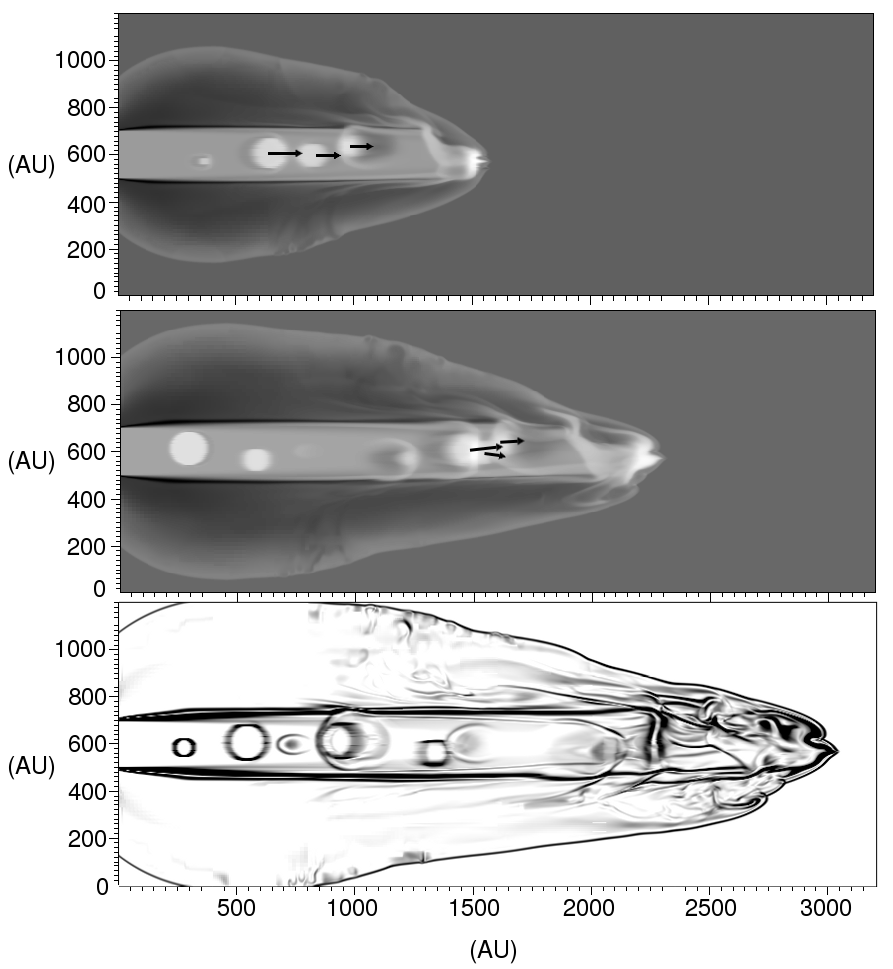}
\caption{The top two panels give grayscale images of logarithmic density in the $x$-$z$ plane located on the jet axis at two different times, $t=$53 \& 77 yr, corresponding to the lower left and upper right panels of Fig. \ref{isocontours}. Lighter gray corresponds to denser material. Velocity vectors originating at three knots have been overlaid, and they are seen to change as the knots interact. The bottom panel shows a synthetic Schlieren image at $t=100$ yr, which illuminates such features as a clump with a forward bow shock at $z=$2100 AU, a clump with reverse bow shock at $z=$900 AU, and clump-induced ``spur shocks'' at several places along the jet beam. The disk-like feature at $z=2300$ AU is discussed further in the text.}
\label{schlieren}
\end{figure}

A time-sequence of the simulation is given over four panels in Fig.~\ref{isocontours}. Density plots and a Schlieren image are shown in Fig.~\ref{schlieren}.  Figure \ref{isocontours} shows a 3-D representation of the simulation in the form of a set of iso-density contours.  In the panels, the jet beam enters from the left hand side of the grid and propagates to the right. Shortly after the start of the simulation, knots appear with random sizes, locations, and speeds. The figure has been adjusted to track the evolution of the clumps via an iso-density contour of a passive clump tracer (in green).  Thus the clumps are readily recognizable as initially spherical inclusions within the beam close to the inflow boundary cells. As the simulation progresses, the clumps evolve via their interaction with the inter-clump material in the beam and, in some cases, with other clumps.
We note that bow shocks form for clumps with high enough differential speeds relative to the beam i.e. $|\Delta v_c| =|v_{c} - v_{j}| > c_j$ where $c_j$ is the sound speed in the jet beam.  As a bow shock propagates into the beam material, a transmitted shock will propagate into and compress the clumps.  If the differential velocity $\delta v_c = \Delta v_c/v_{j}$ is positive then the bow shock faces forward into the direction of jet propagation.  If $\delta v_c$ is negative then the bow shock will face backwards towards the jet source.  These bow shocks are not directly visible in Fig.~\ref{isocontours}, however the sign of $\delta v_c$ is apparent in the compression occurring on the leading (trailing) edges when $\delta v_c$ is positive (negative).  The bow shocks are apparent in Fig. \ref{schlieren}. In the bottom panel, the clump at $z$ = 2100 AU has $\delta v_c > 0$ and hence a forward facing shock forms at its leading edge while the larger clump at $z$ = 900 AU shows the opposite behavior.  We note also the presence ``spur shocks'' in the bottom panel of Fig.~\ref{schlieren}.  These are concave shocks that originate at the edges of the beam and arc away from the beam \citep{heathcote1996hubble}. We find the presence of clumps off-axis naturally leads to the development of such structures.

Once clumps are launched into the jet there are three possible consequences.  The first consequence is the clump propagating downstream unimpeded and colliding with the jet head.  In this case the dense clump may break through the bow shock defining the front edge of the jet, leading to significant non-axisymmetric structures there.  This behavior is apparent in both Fig.'s~\ref{isocontours} and \ref{schlieren} where a dense clump has already traversed the jet length and propagated through the jet shock/bow shock structure at the terminus of the beam.  The presence of a significant ``knob'' protruding at the lower edge of the jet head defines the extent of the clump which now forms the leading edge of the jet.

A second possibility however is that the clump will not make it to the leading edge of the jet.  The behavior of shocked clumps has been extensively studied both analytically and numerically \citep[e.g.][]{klein1994,jones1996,poludnenko2002,fragile2004radclumps}.  These studies demonstrate that a clump compressed by a strong transmitted shock wave (in this case the transmitted wave originates from its relative motion within the beam) will eventually be destroyed in a ``cloud crushing time'' given approximately by $t_{cc} = 2r_c \chi_{cj}^{1/2}/|\Delta v_c|$ where $\chi_{cj}= \rho_c/\rho_j$.  If this happens before the clump reaches the jet head then the clump material will be dispersed within the beam.  Thus the cloud crushing timescale should be compared with the timescale required to cross the length of the jet beam, $t_{jc} = L_j/(v_c-v_{bs})$.  Note that $L_j = v_{bs} t$ where $v_{bs}$ is the speed of the jet head given by the familiar formula $v_{bs} = v_j/ (1+\chi_{ja}^{-1/2})$ and $t$ is the time at which the clump is launched.  Comparing these expressions we find the critical launching time $t^*$,
\begin{equation}
t^* = \frac{2 r_c}{|\Delta v_c|} \sqrt{\chi_{cj}}\left(\frac{v_c}{v_j} (1+\chi_{ja}^{-1/2}) - 1\right)\qquad (v_c\neq v_j)
\end{equation}
A clump of radius $r_c$, velocity dispersion $\Delta v_c$ and density ratio $\chi_{cj}$ needs to be launched at a time $t=t_{launch}<t^*$ in order for it to reach the jet head before being destroyed. For each instantiated clump, comparing $t_{launch}$ to $t^*$ reveals that $t_{launch}<t^*$ for three of the twelve clumps. This implies that late in time (assuming no other interactions) nine of the twelve clumps would disperse before reaching the jet head. However, $t_{cc}>t_{sim}$ for all the clumps in the simulation ($t_{sim}$ is the duration of the simulation). We therefore expect all clumps to exert a strong influence on the jet beam throughout the simulation, and we expect some of the clumps to affect strongly the morphology of the jet head. These expectations are confirmed in the simulation.

The third possibility for the long term evolution of a clump, assuming $\delta v_c$ is not the same for all clumps, is interaction with another clump.  The interaction can take the form of direct or glancing collision depending on the impact parameter $b$.  Even when $b > 2r_c$ there can still be interactions between a clump-driven bow shock and a neighboring clump \citep{poludnenko2002}  Such interactions will be determined by the width of the clump bow shock which will, in general be determined by the Mach number of the clump through the inter-clump beam media ($M_c = |\Delta v_c|/c_j$).  The collisions of clumps is also a process which has been well studied, and one expects the formation of transmitted shock waves within the clumps which heat and compress clump material as it streams into the shock \citep{klein1994, miniati1999}. Figure \ref{isocontours} and Fig.~\ref{schlieren} show a number of such interactions occurring.  By the last panel of Fig.~\ref{isocontours}, clump collisions have resulted in a merged structure near the head of the jet, and their effect on each other and on the jet beam itself is complex. It is noteworthy however that the collision, compression and subsequent merger of clumps can come to resemble the internal working surfaces in homogeneous pulsed jets.  Note the resemblance in the bottom panel of Fig.~\ref{schlieren} of the structure at $z=2300$ AU to the thin shock bounded working surfaces seen in jet pulsation simulations.  An important difference in the case of clumps however is that these structures remain sub-radial. The merged structure identified at $z=2300$ AU has a lateral width of $h \approx 1.8 r_j$ and is offset from the jet axis by $r_0\approx 0.25 r_j$. Thus, even collisions between slow moving clumps overtaken by faster moving ones differ from what is expected for homogeneous jet beams with varying inflow velocity. In this case, the resulting shock structures are slightly smaller than the jet diameter and displaced from the jet axis.

The difference between homogeneity and heterogeneity is particularly striking for glancing clump collisions: $r_c < b < 2r_c$.  In these cases the clump-clump interaction will be off center and one can expect from momentum conservation that non-axial motions will result. The top two panels of Fig.~\ref{schlieren} illustrate this point, showing the off-center collision of three clumps.  Before the collision the velocity vectors of all three clumps is purely axial $\vec{v} = v_z \hat{e}_z$.  After the collision the clumps have acquired transverse $v_r$ velocities.  The ability to generate non-axial motions within the beam via clump interactions is an important point as proper motion studies of highly resolved HH jets show knot to knot variability in both direction and speed \citep{hartigan2005hh47, bally2002hh1/2}. Moreover, this interaction forms the structure at $z=2300$ AU seen in the bottom panel of Fig.~\ref{schlieren} which therefore has varying velocity components across its surface, in contrast to similar structures in pulsed-jet models.
We note again that the clump injection time, position within beam cross-section and velocity were all chosen randomly within constraints. We now address the issue of the observational properties of jets with this kind of heterogeneity, with respect to assumptions of pulsation. In many studies lifetimes of features along an HH beam jet beam are derived by relating the current position of the feature and its proper motion \citep{bally2002hh1/2, hartigan2005hh47}. Going a step further several some studies \citep[e.g.][]{raga1998hh34, raga2002hh34/111}) infer full velocity histories of jets from observations in terms of multiple pulsation modes at the jet launch region.  While these studies are able to convincingly reproduce some observational characteristics of the jets, our models of heterogeneous jets shed new light on the issue of recovering pulsation histories from observations. We note that inferring ejection histories from current positions and velocities requires two intrinsic assumptions: first, that the launching is smoothly varying and periodic in nature, and second, that pulsation-formed knots are coherent throughout their lifetimes. In particular, it is assumed that given a position $z_0$ for a hypersonic, semi-ballistic jet, the time-variability in the past can be inferred from the current observed velocity structure. Considering the current epoch to be $t=0$ one uses observed axial velocity $v(z_0)$ to compute dynamical times $t_0 = -z_0/v(z_0)$ for fluid parcels when they where ejected from the source. The velocity history of the jet as $v_j(t) = v \left[t = -z/v(z)\right]$ can then be derived. In \cite{raga2002hh34/111} for HH~34 and HH~111 it was assumed that $v_j(t)$ could be described by a periodic function of the form
\begin{equation}
v_j(t) = v_{j,0} + \sum_{k=1}^{n} v_j^{(k)} \sin\left(\frac{2\pi t}{\tau_k}+\phi_k\right)
\label{periodic}
\end{equation}
where $ v_j^{(k)}, \tau_k$ and $\phi_k$ are the amplitude, period, and phase of each pulsation mode. For HH~111, $n=1$, while for HH~34, accounting for the collimated knots near the source requires $n=3$.

\begin{figure}[htbp]
\centering
\includegraphics[width=120mm]{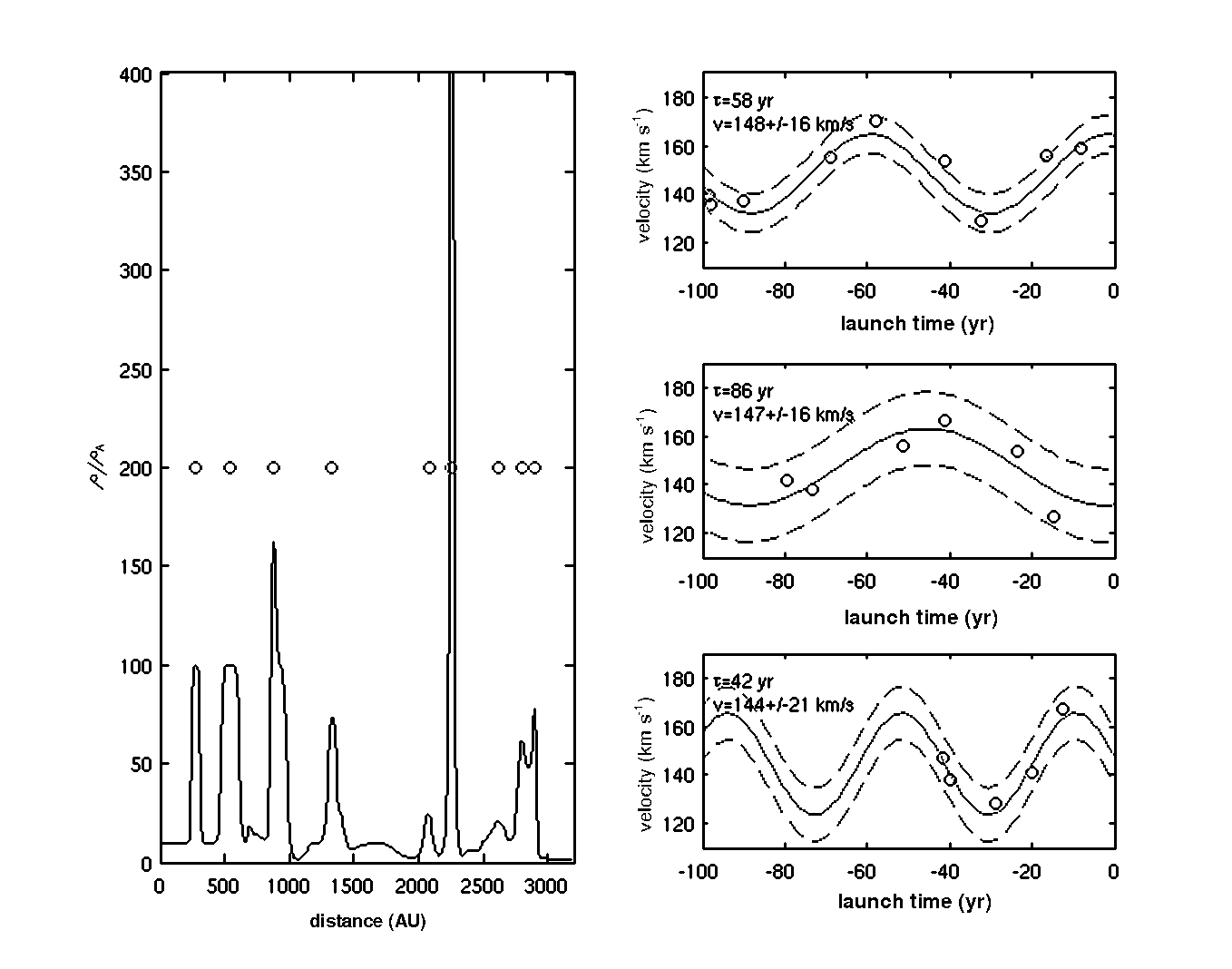}
\caption{$Left\ panel$: The axial density profile normalized by the ambient density, $\rho/\rho_a$, is plotted with positions designated as ``knots'' given by ``$\circ$''. $Right\ panels$: From top to bottom, velocities for knots (``$\circ$'') versus their launching (dynamical) time at $t=$100, 80, \& 40 yr. Results from least-squares fits are shown with 1-$\sigma$ error overlaid. The period $\tau$ and mean velocity with single-mode sinusoidal amplitude, $v_j=v_{j,0}\pm v_j^{(1)}$, are printed on each panel.}
\label{curvefit}
\end{figure}

Assuming Eq.~\ref{periodic} for $v_j(t)$ can bias the description of jets as we now demonstrate. For each data frame in our simulation, we may perform a similar analysis; see Fig.~\ref{curvefit}. Along the axis of the jet we may denote the locations of ``knots'' as those positions where the density peaks at some value above the jet density. The axial density profile is given in the left panel of Fig.~\ref{curvefit}, with knots identified by circles. We then take the velocities at these positions to be the knot velocities. This allows us to calculate dynamical times and, using a single mode ($n=1$) for simplicity, reconstruct a separate velocity history for each data frame in the simulation.
As implied by the second assumption above, in order for this analysis to be valid the resulting amplitude and period should be robust to the time in the simulation, especially late in time when there are the greatest number of knots. We found that the results were not robust, as depicted in the right hand panels of Fig. \ref{curvefit}. The three panels show, from top to bottom, the result of least-squares fits to the data at $t= 100,\ 80,$ and $40$ yr with 1-$\sigma$ error bounds overlaid. The top panel shows the fit resulting from the corresponding data given in the left panel. The average velocity of the knots remains roughly constant for each of the three times depicted and is close to the jet velocity ($v_j = 150$ km s$^{-1}$), as expected. However, the amplitude varies by $\sim30\%$ over time, and the period varies by over $100\%$. The goodness of the fits, though less at $t=80$ yr than the other two times depicted, appears adequate. Although not explicitly shown here, variation in the fit results shows correlation with knot-knot interactions.

\section{Discussion and Conclusions}
We have performed 3-D simulations using a new ``collimated clumps'' scenario for protostellar jets.  Our model offers a fundamentally different paradigm for understanding jet origins and dynamics in the sense that heterogeneity is seen as being intrinsic and links the jet morphology on ``meso-scales'' to the processes (such as instabilities) occurring on ``micro-scales'' near the central engine.

While the pulsed-jet model has been successful at interpreting some aspects of jets, it may be misleading if used too generally. In particular, the assumption of sinusoidal pulsations can limit the interpretation of HH object observations. In our simulations, which had no sinusoidal variation in time, we nonetheless were able to recover (erroneous) sinusoidal behavior using an analysis similar to that which has been carried out in the past on observations. While one part of this behavior (the mean velocity $v_{j,0}$) fit the initial conditions well, the assumption of periodic pulsation allows the false conclusion that the structures in the beam arose due to periodic ejection behavior. We conclude therefore that care should be taken when attributing observations of apparent sinusoidal velocity in protostellar jets to corresponding sinusoidal behavior of a central engine.

In contrast to pulsed-jet models, our model offers two attractive features: first, a natural mechanism (knot-knot interactions) helps explain small-scale features along the jet axis. The idea that knots or bow shocks in HH objects are evolving clumpy structures has been discussed before in the context of observations \citep[e.g.][]{reipurth2002hh34}, with ongoing HST observations giving continued support to this idea \citep[e.g.][]{hartigan2005hh47}. The interaction of distinct sub-radial knots with each other and with an overall bow shock offers a simple explanation for such evolution. The second feature the model offers is the presence of unique observational characteristics in the form of forward- and reverse-facing bow shocks and spur shocks at the edges of the jet beam. Our scenario provides a simple mechanism for the formation of sub-radial non-axisymetric features via multiple dense, clumps which have a nonzero velocity dispersion. The ease with which non-axisymmetric features like spur shocks \citep{heathcote1996hubble} develop in our models is attractive.

What would be the origin of the entrained clumps? The issues of the stability of jets particularly at ``micro-scale'' regions near the launch region remains unresolved \citep{konigl2004diskwinds, ouyed2003jetstability, xu2000jetstability3d, micono1998khjets}. In addition, issues of magnetic field strength and velocity perturbations near the jet launch region also present problems \citep{hartigan2007bfields}. Although we do not explicitly consider details of disk/jet launching here, plausible scenarios present themselves. It has long been considered that the interaction between the forming star and accretion disk---via magnetic fields---is responsible for jet launching \citep{konigl1982}. Further, anisotropy of the magnetized disk during the YSO phase is a distinct possibility, as suggested by \cite{combet2008}. One could therefore theorize the occurrence of a non-cylindrically-symmetric accretion burst as a possible explanation for non-axial density enhancements in the jet beam.

Recent high energy density laboratory astrophysics (HEDLA) investigations provide an unique window into the behavior of fully 3-D radiative hypersonic MHD jets.  These experiments demonstrate that magnetized jet beams in the lab may break up into a sequence of quasi-periodic knots due to the kink instability \citep{ciardi2007labjet}. These knots may be displaced slightly to the side of the nominal jet axis and may propagate with varying velocities. This results in morphologies qualitatively reminiscent of HH-jet beams. It should be noted that the present simulation does not employ magnetic fields; however, it remains an open question whether magnetic fields remain dynamically important on the length scales of consideration here \citep{hartigan2007bfields, ostriker2001molecularclouds}. It seems plausible that a similar process could occur in the astrophysical context, beginning with a smooth beam near the central engine which then becomes disrupted owing to the kink or sausage instabilities on intermediate scales. This would result in a series of knots which continue to evolve as they propagate away from the central engine. Such a scenario would also explain the observed velocity differences between knots, attributable to the particulars of each knot's formation. The present simulation is an idealization of this model.

The degree to which the results of the simulation presented here depend on the values chosen for parameters such as $\chi_{cj}$ and $\Delta v_c$ requires further investigation. In particular, $t_{launch}<t^*$ was not satisfied for all of the instantiated clumps, implying that some should disperse before reaching the jet head. The present simulation did not progress far enough to witness this behavior (that is, $t_{cc}>t_{sim}$). Also, using a $\Delta v_c\neq0$ and varying from clump to clump should result in an injection of considerable vorticity into the jet beam. Future work should therefore explore the long term evolution of the clumped jet. The aspect ratio (length/width) of the jet's final state in the present simulation is less than that for HH objects, implying again that simulations which progress for longer times would be of interest. While increased resolution would of course be of benefit, extracting details of clump-clump interaction within the jet-beam environment is probably outside the scope of this study and would be appropriate for a separate investigation (Dennis et al. 2008, in preparation). Finally, the inclusion of magnetic fields to study scenarios of knot formation and the resulting observational consequences would be of additional interest.

\acknowledgements
We would like to thank Pat Hartigan, Sergey Lebedev, and Andrea Ciardi for their time. Tim Dennis and Brandon Shroyer also provided invaluable support and help.

Support for this work was in part provided by by NASA through awards issued by JPL/Caltech through Spitzer program 20269 and 051080-001, the National Science Foundation through grants AST-0507519 as well as the Space Telescope Science Institute through grants HST-AR-10972, HST-AR-11250, HST-AR-11252 to.   We also thank the University of Rochester
Laboratory for Laser Energetics and funds received through the DOE Cooperative Agreement No. DE-FC03-02NA00057.

\bibliography{yirak}

\end{document}